\documentclass[twocolumn,aps,prb,showpacs,preprintnumbers,amsmath,amssymb, superscriptaddress]{revtex4-1}
\usepackage{bm}
\usepackage{graphicx}
\usepackage{color,xspace}
\usepackage{ulem}
\usepackage{float}

\newenvironment{addendum}{%
   \setlength{\parindent}{0in}%
   \small%
   \begin{list}{Acknowledgements}{%
       \setlength{\leftmargin}{0in}%
       \setlength{\listparindent}{0in}%
       \setlength{\labelsep}{0em}%
       \setlength{\labelwidth}{0in}%
       \setlength{\itemsep}{12pt}%
       }
   }
   {\end{list}\normalsize}

\begin{document}

\title{Multiple topological states in iron-based superconductors}

\author{Peng~Zhang}
\affiliation{Institute for Solid State Physics, University of Tokyo, Kashiwa, Chiba 277-8581, Japan}

\author{Zhijun Wang}
\affiliation{Department of Physics, Princeton University, Princeton, New Jersey 08544, USA}

\author{Xianxin Wu}
\affiliation{Institut f\"ur Theoretische Physik und Astrophysik, Julius-Maximilians-Universit\"at W\"urzburg, W\"urzburg, 97074, Germany}

\author{Koichiro Yaji}
\affiliation{Institute for Solid State Physics, University of Tokyo, Kashiwa, Chiba 277-8581, Japan}

\author{Yukiaki Ishida}
\affiliation{Institute for Solid State Physics, University of Tokyo, Kashiwa, Chiba 277-8581, Japan}

\author{Yoshimitsu Kohama}
\affiliation{Institute for Solid State Physics, University of Tokyo, Kashiwa, Chiba 277-8581, Japan}

\author{Guangyang Dai}
\affiliation{Beijing National Laboratory for Condensed Matter Physics and Institute of Physics, Chinese Academy of Sciences, Beijing 100190, China}

\author{Yue Sun}
\affiliation{Institute for Solid State Physics, University of Tokyo, Kashiwa, Chiba 277-8581, Japan}
\affiliation{Department of Applied Physics, University of Tokyo, Bunkyo-ku, Tokyo 113-8656, Japan}

\author{Cedric Bareille}
\affiliation{Institute for Solid State Physics, University of Tokyo, Kashiwa, Chiba 277-8581, Japan}

\author{Kenta Kuroda}
\affiliation{Institute for Solid State Physics, University of Tokyo, Kashiwa, Chiba 277-8581, Japan}

\author{Takeshi Kondo}
\affiliation{Institute for Solid State Physics, University of Tokyo, Kashiwa, Chiba 277-8581, Japan}

\author{Kozo Okazaki}
\affiliation{Institute for Solid State Physics, University of Tokyo, Kashiwa, Chiba 277-8581, Japan}

\author{Koichi Kindo}
\affiliation{Institute for Solid State Physics, University of Tokyo, Kashiwa, Chiba 277-8581, Japan}

\author{Xiancheng Wang}
\affiliation{Beijing National Laboratory for Condensed Matter Physics and Institute of Physics, Chinese Academy of Sciences, Beijing 100190, China}

\author{Changqing Jin}
\affiliation{Beijing National Laboratory for Condensed Matter Physics and Institute of Physics, Chinese Academy of Sciences, Beijing 100190, China}

\author{Jiangping Hu}
\affiliation{Beijing National Laboratory for Condensed Matter Physics and Institute of Physics, Chinese Academy of Sciences, Beijing 100190, China}

\author{Ronny Thomale}
\affiliation{Institut f\"ur Theoretische Physik und Astrophysik, Julius-Maximilians-Universit\"at W\"urzburg, W\"urzburg, 97074, Germany}

\author{Kazuki Sumida}
\affiliation{Graduate School of Science, Hiroshima University, Higashi-Hiroshima 739-8526, Japan}

\author{Shilong Wu}
\affiliation{Hiroshima Synchrotron Radiation Center, Hiroshima University, Higashi-Hiroshima 739-0046, Japan}

\author{Koji Miyamoto}
\affiliation{Hiroshima Synchrotron Radiation Center, Hiroshima University, Higashi-Hiroshima 739-0046, Japan}

\author{Taichi Okuda}
\affiliation{Hiroshima Synchrotron Radiation Center, Hiroshima University, Higashi-Hiroshima 739-0046, Japan}

\author{Hong Ding}
\affiliation{Beijing National Laboratory for Condensed Matter Physics and Institute of Physics, Chinese Academy of Sciences, Beijing 100190, China}

\author{G.D. Gu}
\affiliation{Condensed Matter Physics and Materials Science Department, Brookhaven National Laboratory, Upton, NY 11973, USA}

\author{Tsuyoshi Tamegai}
\affiliation{Department of Applied Physics, University of Tokyo, Bunkyo-ku, Tokyo 113-8656, Japan}

\author{Takuto Kawakami}
\affiliation{Yukawa Institute for Theoretical Physics, Kyoto University, Kyoto 606-8502, Japan}

\author{Masatoshi Sato}
\affiliation{Yukawa Institute for Theoretical Physics, Kyoto University, Kyoto 606-8502, Japan}

\author{Shik Shin}
\affiliation{Institute for Solid State Physics, University of Tokyo, Kashiwa, Chiba 277-8581, Japan}

\date{\today}

\maketitle

\textbf{Topological insulators and semimetals as well as unconventional iron-based superconductors have attracted major recent attention in condensed matter physics. Previously, however, little overlap has been identified between these two vibrant fields, even though the principal combination of topological bands and superconductivity promises exotic unprecedented avenues of superconducting states and Majorana bound states (MBSs), the central building block for topological quantum computation. Along with progressing laser-based spin-resolved and angle-resolved photoemission spectroscopy (ARPES) towards high energy and momentum resolution, we have resolved topological insulator (TI) and topological Dirac semimetal (TDS) bands near the Fermi level ($E_{\text{F}}$) in the iron-based superconductors Li(Fe,Co)As and Fe(Te,Se), respectively. The TI and TDS bands can be individually tuned to locate close to $E_{\text{F}}$ by carrier doping, allowing to potentially access a plethora of different superconducting topological states in the same material. Our results reveal the generic coexistence of superconductivity and multiple topological states in iron-based superconductors, rendering these materials a promising platform for high-$T_{\text{c}}$ topological superconductivity. }

High-$T_{\text{c}}$ iron-based superconductors feature multiple bands near $E_{\text{F}}$, which enhances the difficulty in understanding the details of unconventional pairing \cite{HosonoJACS2008, JohnstonAIP2010, StewartRMP2011}. It, however, also allows for a wealth of, possibly topologically non-trivial, electronic bands, of which a recent example is the TI states discovered in the iron-based superconductor Fe(Te,Se)~\cite{ZhangScience2018}, hinting at a promising direction to realize topological superconductivity and MBSs \cite{FuPRL2008, MourikScience2012, YazdaniScience2014, AlbrechtNature2016, MelePRL2013}. In view of Fe(Te,Se), a pressing subsequent  question is to which extent this marks a generic phenomenon in different classes of iron-based high-$T_{\text{c}}$ superconductors.
In this work, we find that the emergence of non-trivial topological bands near the Fermi level is indeed a common feature of various iron-based superconductors. Our first-principles calculations reveal that BaFe$_2$As$_2$, LiFeAs and Fe(Te,Se) all exhibit band inversions along $k_z$. To confirm these calculations, the band structures of Li(Fe,Co)As and Fe(Te,Se) were investigated by laser-based high-resolution ARPES. Firstly, we observe that TI bands reminiscent of Fe(Te,Se) exist in Li(Fe,Co)As as well, supporting the generic existence of non-trivial topology in iron-based superconductors. Secondly and more interestingly, we predict and observe TDS bands in Li(Fe,Co)As and Fe(Te,Se), which we investigate via high-resolution ARPES, spin-resolved ARPES (SARPES), and magnetoresistance (MR) measurements. Finally, we discuss the phase diagram of these topological high-$T_{\text{c}}$ compounds as a function of Fermi level (doping). The combination of topological states and superconductivity may produce not only surface topological superconductivity deriving from the TI edge states, but also bulk topological superconductivity from the TDS bands.

\begin{figure*}[!htb]
\begin{center}
\includegraphics[width=0.9\textwidth]{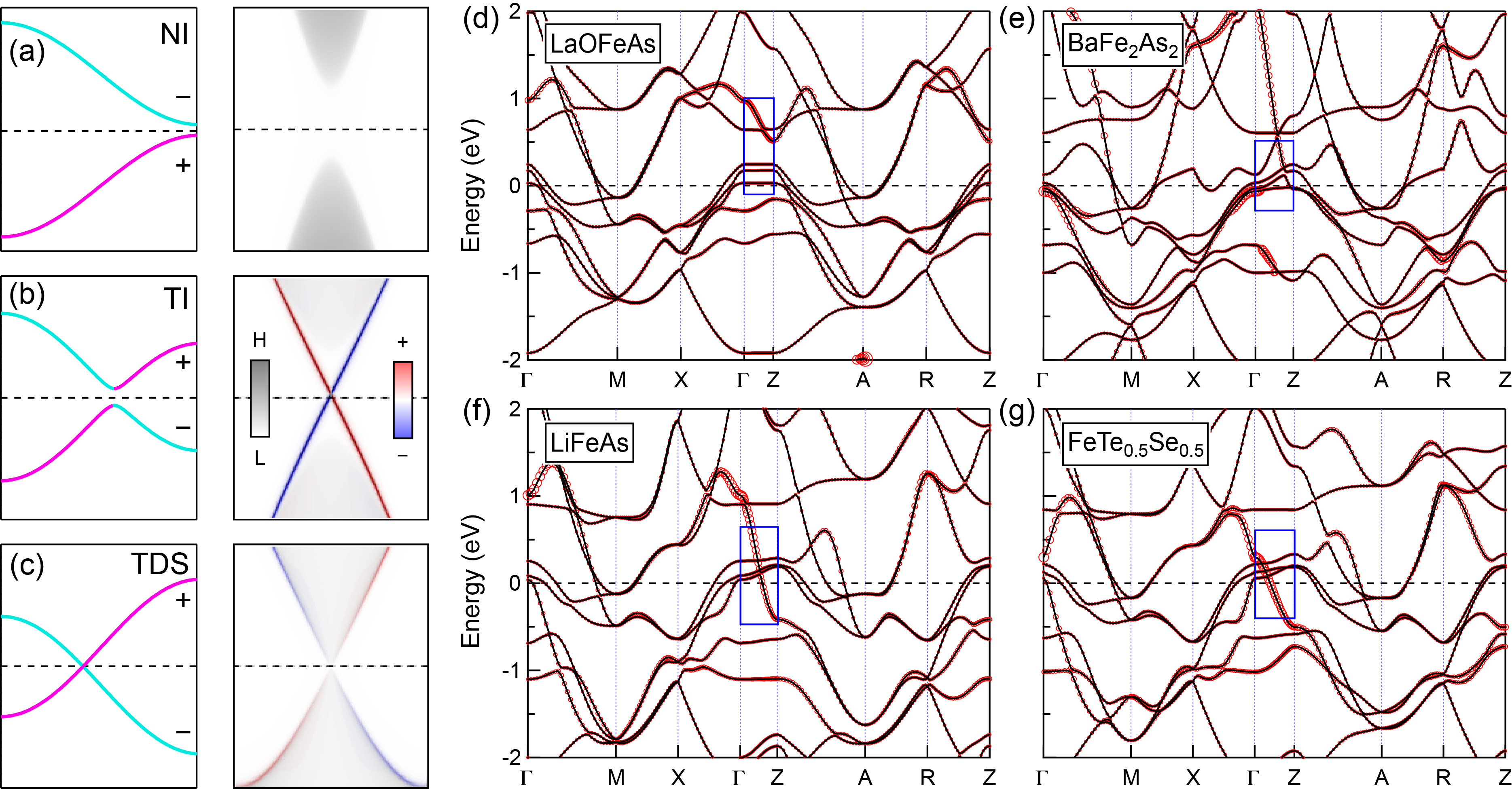}
\end{center}
 \caption{\label{band}  \textbf{Different topological phases and band structures of iron-based superconductors.} (a - c) The out-of-plane dispersion and the corresponding (001) surface spectrum, for NI (a), TI (b) and TDS (c). We overlapped the spin-polarized surface spectrum (blue and red) on top of the spin-integrated surface spectrum (grey). (d - g) Band structures of LaOFeAs, BaFe$_2$As$_2$, LiFeAs and FeTe$_{0.5}$Se$_{0.5}$, respectively. The size of red circles represent the weights of As/Se $p_z$ orbital. No band inversion between $p_z$ and $d_{xz}$/$d_{yz}$ is found in LaOFeAs, while there are band inversions in BaFe$_2$As$_2$, LiFeAs and FeTe$_{0.5}$Se$_{0.5}$.}
\end{figure*}

Normal insulator (NI), TI, and TDS constitute topologically distinct phases\cite{HasanRMP2010, QiRMP2011, FangPRB2012, FangPRB2013}. From a band structure point of view, if there is no band inversion in whole BZ, the material will be an NI, with no spin-polarized surface states, as displayed in Fig. 1(a). In a simple case, if there is a single band inversion in the whole BZ, the material will be topologically non-trivial, with spin-polarized Dirac-cone type bands. In a TI, the bulk band gap leads to well-defined surface states, with explicit spin helicity [Fig. 1(b)]. Instead, in a TDS,  the band crossing is protected by the crystal symmetry, and the surface Dirac  band generally overlap with the bulk Dirac band on the (001) surface \cite{FangPRB2012, FangPRB2013, HasanNC2015, LanzaraNC2016}. Without bulk band gap, the spin helicity of the surface states is not fixed (Supplementary Information Part IV). The spin-polarization magnitude of the (001) surface states generally show a gradual increase with the increased distance from the Dirac point, as shown in Fig. 1(c). The three phases may coexist in one material.

In the previous papers, we show that there is a band inversion between the $p_z$ and $d_{xz}$/$d_{yz}$ bands in Fe(Te,Se), resulting in TI states \cite{WangPRB2015, ZhangScience2018}. Here we further check the band structure of the four major classes of iron-based superconductors, focusing on the band inversions of $p_z$ and $d_{xz}$/$d_{yz}$ bands along $k_z$. The separations between the adjacent FeAs/FeSe layers $\Delta_d$ are 8.741, 6.508, 6.364 and 5.955 \AA~for LaOFeAs (1111), BaFe$_2$As$_2$ (122), LiFeAs (111) and FeTe$_{0.5}$Se$_{0.5}$ (11), respectively \cite{JohnstonAIP2010, WangPRB2015}.
This separation directly determines the interlayer $pp$ coupling, which will affect the band width of the $p_z$ band along $\Gamma$Z \cite{WangPRB2015}. The parameter $\Delta_d$ for LaOFeAs is much larger than BaFe$_2$As$_2$, LiFeAs and Fe(Te,Se), resulting in a small $p_z$ dispersion for LaOFeAs but large $p_z$ dispersions for BaFe$_2$As$_2$, LiFeAs and Fe(Te,Se). In Fig.1(d - g), we display the band structures of the four classes. It is clear that along the $\Gamma$Z direction there is no band inversion for LaOFeAs, while there are band inversions for BaFe$_2$As$_2$, LiFeAs and Fe(Te,Se). In 2D thin films, such as Fe(Te,Se) film, there is no $k_z$ dispersion. However, the band inversion may still happen for the in-plane band structure. The in-plane lattice parameter $a$ (or Se/Te height) affects the intralayer $pd$ coupling and determine the position of $p_z$ band at $\Gamma$. By reducing the parameter $a$ (increasing the Se/Te height), $p_z$ band will sink below $d_{xz/yz}$ band at $\Gamma$, generating a band inversion and realizing a 2D topological insulator \cite{WuPRB2016, DingSB2017}

\begin{figure*}[!htbp]
\begin{center}
\includegraphics[width=0.85\textwidth]{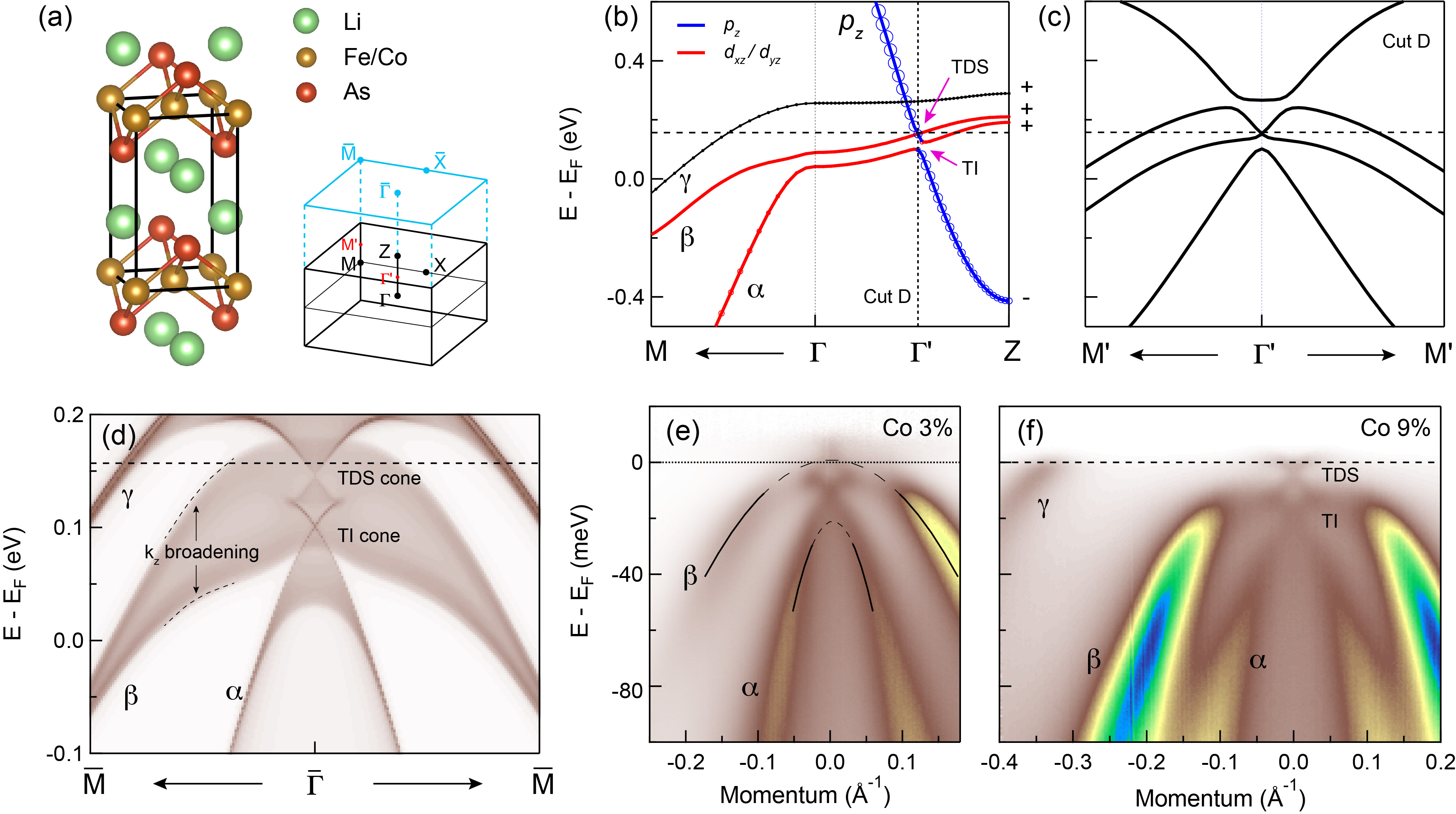}
\end{center}
 \caption{\label{spin} \textbf{Electronic structure of Li(Fe,Co)As. } (a) Left: Crystal structure of Li(Fe,Co)As.  Right: High symmetry points in the 3D BZ and (001) surface BZ. (b) Zoomed-in view of the LiFeAs band dispersion along $\Gamma$M and $\Gamma$Z. The red color marks the $d_{xz}$/$d_{yz}$ bands, and the blue color marks the $p_z$ band. The marker size relates to the weight of the $p_z$ orbital character. The avoided crossing between the $p_z$ and $\alpha$ bands produces TI states, while the real crossing between the $p_z$ and $\beta$ bands produces TDS states. The crossing between the $p_z$ and $\gamma$ bands also produces TDS states far above $E_{\text{F}}$. (c) In-plane band structure at Cut D, where the bulk Dirac cone of the TDS bands is shown. (d) (001) surface spectrum of LiFeAs. There is a large broadening in the spectrum at the $\alpha$ and $\beta$ band tops with reduced intensity, due to the $k_z$ dispersion. (e) ARPES intensity plot of LiFe$_{1-x}$Co$_x$As ($x$ = 3\%) at 15 K, with $p$-polarized photons. The spectrum is divided by the corresponding Fermi function. (f) ARPES intensity plot of LiFe$_{1-x}$Co$_x$As ($x$ = 9\%) at 10 K, with $p$-polarized photons. The two Dirac cones are similar to the ones in the calculation (d). }
\end{figure*}

Band structure measurements are required to verify the existence of the non-trivial topology. Since LiFeAs has no magnetic and structural transitions compared to BaFe$_2$As$_2$, we carried out a detailed study on the band structure of Li(Fe,Co)As \cite{JinSSC2008, BorisenkoPRL2010, PitcherJACS2010, PlattPRB2011, MiaoPRB2014, MiaoNC2015}.
There are reports on the band inversion in LiFeAs and NaFeAs \cite{WangPRB2015, WatsonPRB2018}. But the related topological bands are still unclear due to the low resolutions. The laser-based high-resolution ARPES and SARPES make it possible to directly resolve the topological bands.
The crystal structure of Li(Fe,Co)As is shown in Fig.2(a). Since the Li atoms are quite close to the FeAs plane, the lattice parameter $c$ is comparable to that of Fe(Te,Se). Thus a large $p_z$ dispersion is also present in LiFeAs, as shown in the zoomed-in view of the band structure along $k_z$ in Fig.2(b). 
The spin-orbit coupling (SOC) splits the $d_{xz}$/$d_{yz}$ bands, forming two hybridized bands $\alpha$ and $\beta$, which both have mixed $d_{xz}$/$d_{yz}$ orbitals along $k_z$. When the $p_z$ band with odd parity (``$-$'') crosses the $\alpha$ band with even parity (``+''), band inversion is formed. The SOC produces an avoided crossing between the $p_z$ and $\alpha$ bands, resulting in TI states, similar to that of Fe(Te,Se)~\cite {ZhangScience2018}. We further notice that the $\beta$ band also has an even parity (``+''). The crossing between the $p_z$ and $\beta$ bands is protected by the crystal $C_4$ rotation symmetry, and forms a 3D Dirac cone. Consequently, the band inversion and the protected band crossing produce TDS states. The bulk Dirac cone of the TDS bands can be seen from the in-plane band structure at Cut D, as shown in Fig.2(c). The surface Dirac cones of both TI and TDS bands can be seen from the (001) surface spectrum in Fig.2(d).
Since bulk bands have $k_z$ dispersions and surface bands do not, in the surface spectrum the bulk bands generally appear as broad continuums and the surface bands appear as sharp features. In Fig.2(d), The $\alpha$ and $\beta$ band tops and whole $p_z$ band appear as broad and weak continuums due to their $k_z$ dispersions, while the the two surface Dirac cones of the TI and TDS bands are very sharp.

By changing Fermi level with different Co content, we observed both TI and TDS surface Dirac cones in Li(Fe,Co)As at the same time. The ARPES band structure of Li(Fe,Co)As with 3\% Co is displayed in Fig.2(e). Despite the surface Dirac cone of the TI bands, the second surface Dirac cone of the TDS bands, which is above $E_{\text{F}}$, shows up in the ARPES spectrum divided by the corresponding Fermi function.
We further checked Li(Fe,Co)As sample with 9\% Co, and displayed in Fig.2(f). As expected, the TDS surface cone shifts down and the full cone clearly shows up, directly confirming the existence of the TDS states.
The bulk band tops of $\alpha$ and $\beta$ bands are not visible neither in Fig.2(e) nor Fig.2(f), similar to the calculated surface spectrum in Fig.2(d), which means that the 7-eV laser-ARPES spectrum of Li(Fe,Co)As is better described by the surface spectrum, rather than the bulk bands at a specific $k_z$.

\begin{figure*}[!htb]
\begin{center}
\includegraphics[width=.9\textwidth]{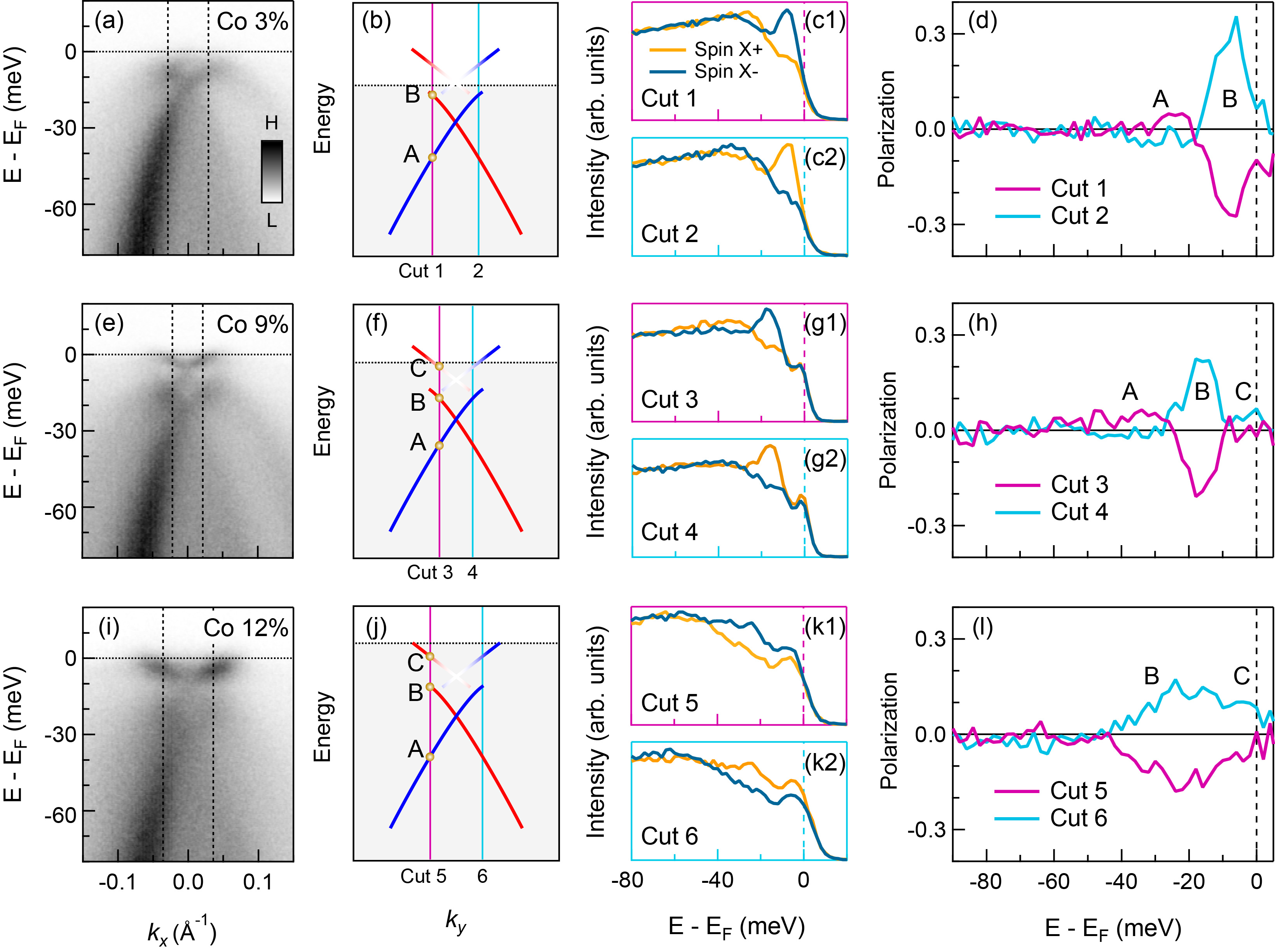}
\end{center}
 \caption{\label{theory} \textbf{Spin polarization of the surface Dirac cones from the TI and TDS states in Li(Fe$_{1-x}$Co$_x$)As.} (a) Intensity plot along $k_x$ for $x$ = 3\% sample, with $p$-polarized photons. (b) Sketch of the spin-polarized surface bands along $k_y$. The blue and red colors stand for opposite spin polarizations, as confirmed by the data in (c  - d). We measured spin resolved EDCs at the cuts indicated by pink (Cut 1) and cyan (Cut 2) lines. The two lines are also duplicated in (a) as dashed lines. Note that the two cuts in (b) are at different $k_y$, which is to avoid the intensity asymmetry in (a) (Supplementary Information Part I). Because of the $C_4$ symmetry, the band structures along $k_y$ and along $k_x$ are the same, despite the intensity difference. (c) Spin-resolved EDCs at Cut 1 (c1) and Cut 2 (c2), respectively. (d) Spin polarization at Cut 1 and Cut 2. The spin along $x$ direction is measured. A and B correspond to the positions indicated in (b). (e - h) Same as (a - d), but for samples with $x$ = 9\%. (i - l) same as (a - d), but for samples with $x$ = 12\%. }
\end{figure*}

As we show in Fig.1(b,c), both TI and TDS surface Dirac cones should be spin polarized. Thus we can use SARPES~\cite{YajiRSI2016} to double check their surface nature. As shown in Fig.3, we measured three different compositions of LiFe$_{1-x}$Co$_x$As with SARPES. The spin-integrated band structures are the same as the ones in Fig.2, except some intensity difference induced by the experimental geometries (Supplementary Information Part I). A pair of spin-polarized EDCs along $k_y$ was measured, with the EDC positions illustrated in Fig.3(b, f, j).
We first focus on the spin polarization of the TI Dirac cone, which is obtained from both $x$ = 3\% and $x$ = 9\% samples, and shown in Fig.3(a - d) and (e - h). In both samples, the lower part of the cone (position A) shows opposite spin polarization with the upper part (position B), and the left part of the cone (Cut 1 and Cut 3) shows opposite spin polarization with the right part (Cut 2 and Cut 4), as expected for the spin-polarization from a Dirac cone. The direction of the spin polarization indicates that the Dirac cone has a left-hand helicity, the same as that of Fe(Te,Se) and most TIs.
As we discussed in Fig.1, the TDS surface Dirac cone is similar to that of TI, whose spin polarization can be obtained from both $x$ = 9\% and $x$ = 12\% samples, as shown in Fig.3(e - h) and (i - l). Since the lower part of the TDS surface cone shows very weak intensity [Fig.3(e, i)], we focus on the spin polarization of the upper part (position C). In the $x$ = 9\% sample [Fig.3(g - h)], a weak spin-polarization at position C is resolved. In the $x$ = 12\% sample [Fig.3(k - l)], a pair of EDCs more distant from the conal point is measured and a larger spin polarization is observed. These results are consistent with the model calculations in Fig.1(c): The spin polarization magnitude of the TDS surface cone is smaller near the conal point, while larger far from the conal point. The experimental spin polarizations of the TI and TDS surface cones are summarized in Fig.3 (b, f, j). The spin polarizations confirm that the two Dirac cones indeed come from the surface states.

\begin{figure}[!htb]
\begin{center}
\includegraphics[width=0.5\textwidth]{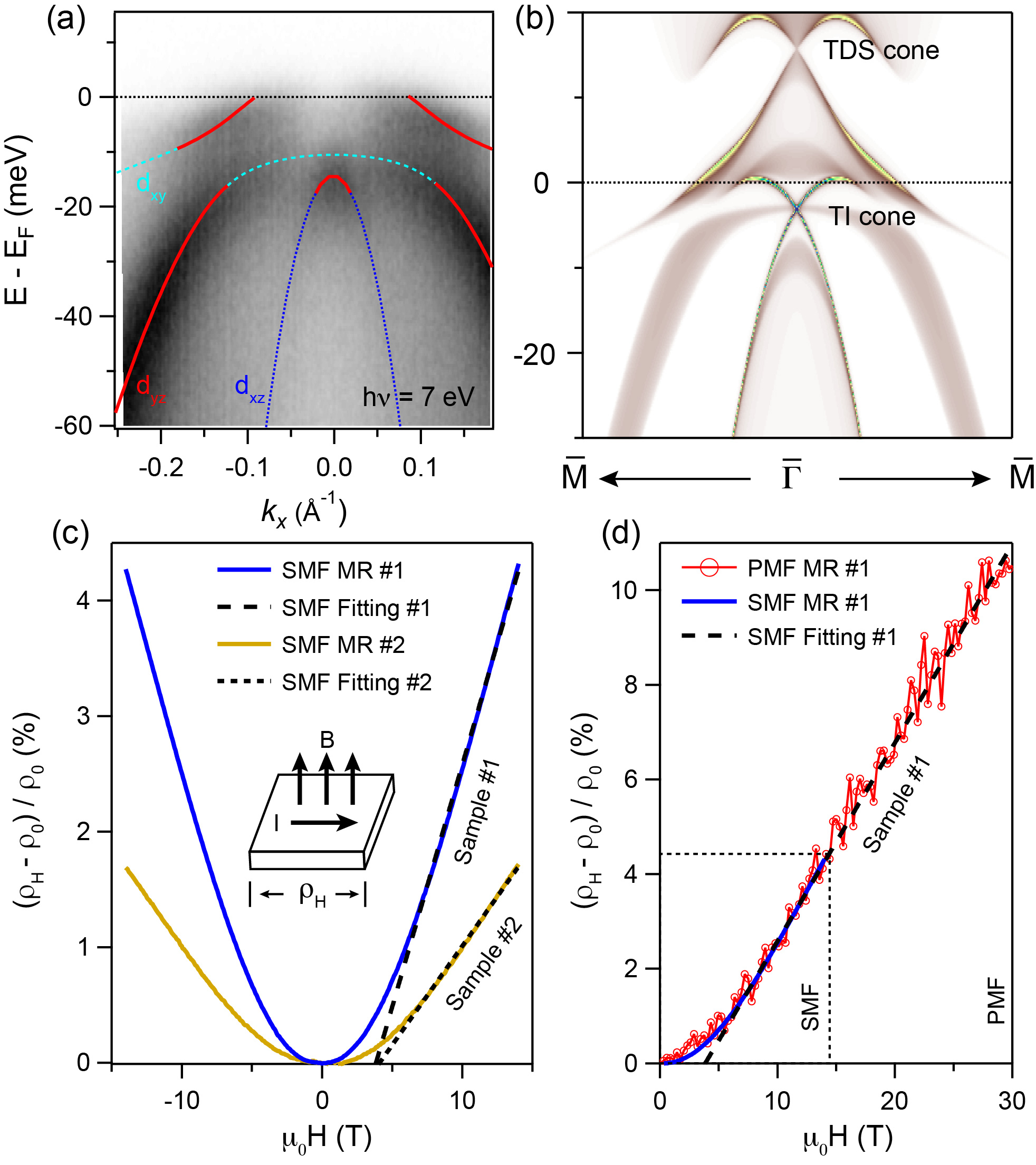}
\end{center}
 \caption{\label{mr} \textbf{The TDS bands in Fe(Te,Se) and the linear MR.} (a) ARPES intensity plot of band structure measured with a laser delivering $s$-polarized 7-eV photons. (b) Calculated (001) surface spectrum based on the $k \cdot p$ model. (c) MR measured on two different Fe(Te,Se) samples in a SMF up to 14 T. Solid curves are the experimental results, while dashed lines are linear fittings of the MR curves in the range of 6 - 14 T. The different MR values of the two samples may come from the magnetic scattering by different content of excess Fe~\cite{SunPRB2014}.  (d) MR measured in a PMF up to 30 T. The red curve represents the data from the PMF measurements. The blue solid curve and black dashed line are duplicates of the SMF results on sample \#1 in (c). Both MR experiments were carried out at 16 K.}
\end{figure}

Because of the similar band structure of Li(Fe,Co)As and Fe(Te,Se), the TDS states may also exist in Fe(Te,Se). The detailed band structure of Fe(Te,Se) from high-resolution ARPES is displayed in Fig.4(a) and Supplementary Information Part II and III. The $d_{xy}$ orbital is below $d_{yz}$ orbital at the $\Gamma$ point, and a hybridization gap opens at their crossing points, similar to the case of FeSe~\cite{ColdeaPRB2015, ZhangPRB2015}, but slightly different from the DFT calculations. Thus, a $k \cdot p$ model based on the real band structure is built to describe the topological states more accurately (Supplementary Information Part III). The TDS bands are clearly shown in the (001) surface spectrum in Fig.4(b). Since there is no mixing of the TDS states and other bulk states near $\Gamma$, evidences of the TDS states may appear in the transport measurements.
It is well known that Dirac and Weyl semimetals, which host bulk Dirac bands, generically show an MR that is linearly dependent on the magnetic field~\cite{FangPRB2013, OngNM2014, LuPRB2015}, which can be explained by the quantum MR~\cite{AbrikosovPRB1998}.
If there are TDS bands near $E_{\text{F}}$ in Fe(Te,Se), it is very likely that such a linear MR should also be realized, the measurement of which we report in the following.
The MR was measured at 16K on two batches of samples (samples \#1 and \#2) with different growing methods (see Methods). Both samples show similar MR curves in a static magnetic filed (SMF), as shown in Fig.~4(c). Indeed, the MR curve above 6 T shows a quantum linear behavior, while the curve below 6 T exhibits a semiclassical quadratic dispersion. The linear fitting in the range of 6 -14 T matches well with the experimental curve.
We also checked the MR in pulsed high magnetic field (PMF) up to 30 T on sample \#1, and display the results in Fig.~4(d). The MR in PMF in the range of 0 - 14 T is the same as that measured in SMF. Above 14 T, the PMF MR exactly follows the extrapolation of linear fitting of SMF MR. All these results clearly show the existence of the linear MR above 6 T in Fe(Te,Se).
We note that there are reports of topologically trivial bulk Dirac-bands near the M point in magnetic BaFe$_2$As$_2$~\cite{RichardPRL2010} or nematic FeSe~\cite{FengPRB2016}. In contrast, however, in Fe(Te,Se) there is no report on such orders, and no Dirac cone was observed~\cite{DingPRB2012, KanigelSA2017}.
Thus, this linear MR most likely comes from the TDS bands.

\begin{figure*}[!htbp]
\begin{center}
\includegraphics[width=0.9\textwidth]{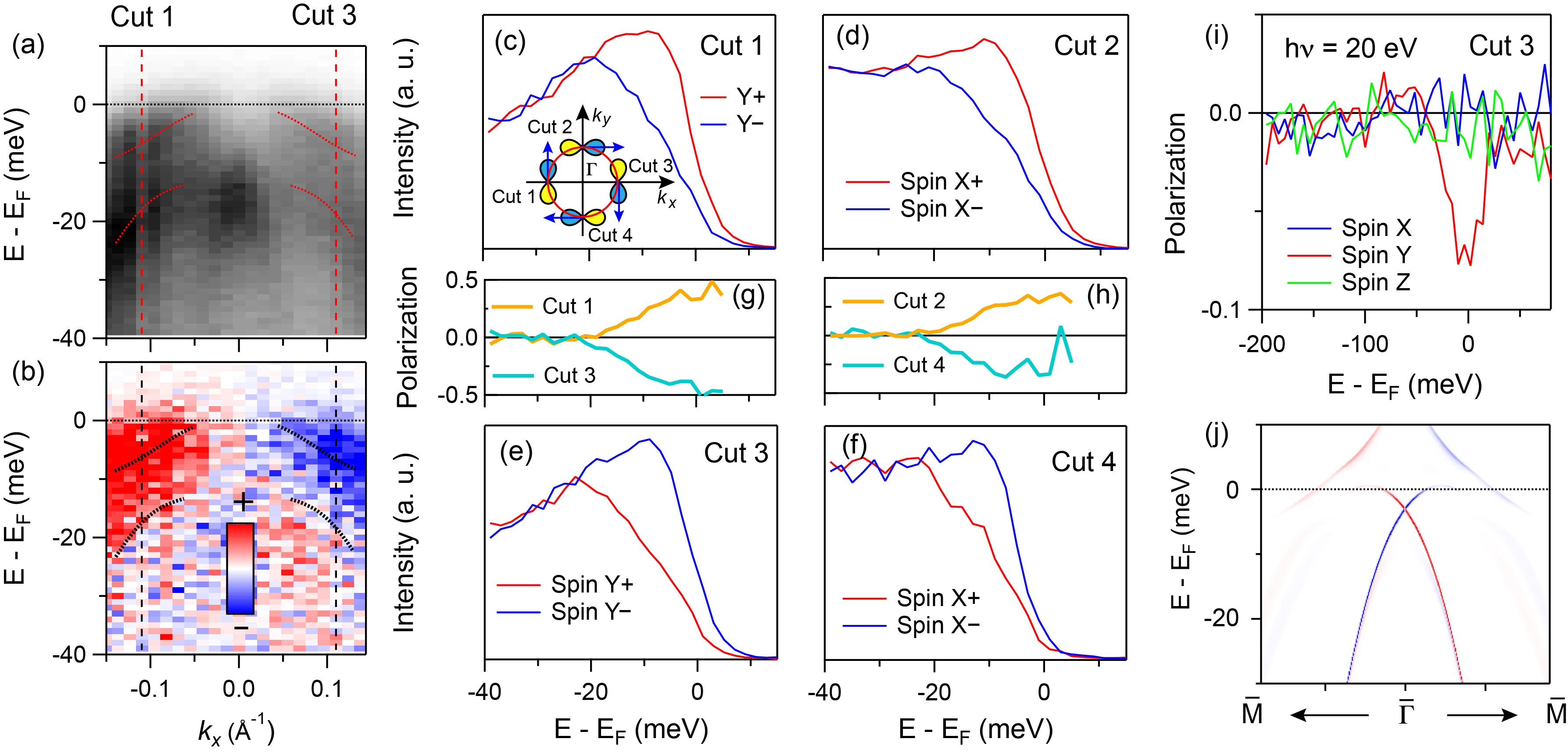}
\end{center}
 \caption{\label{spin} \textbf{Spin polarization of the $d_{yz}$ band.} (a) - (b)  Spin-resolved intensity plot at $\Gamma$ with $s$-polarized 7-eV photons. The intensity scales the sum (a) or difference (b) of spin-up and spin-down photoelectrons along the $y$ direction. (c) - (f) Spin-resolved EDCs at Cut 1-4. Cut 1 and 3 are taken with $s$-polarized photons, while Cut 2 and 4 are taken with $p$-polarized photons, due to the matrix element effect. The inset of (c) shows the positions of the four cuts, and the spin directions are extracted from (g - h). (g - h) Spin-polarization curves at Cut 1-4. (i) The spin polarization along the $x$, $y$ and $z$ directions at Cut 3 measured with 20-eV photons from synchrotron radiation. (j) Spin-resolved (001) surface spectrum calculated from the $k\cdot p$ model.  }
\end{figure*}

We may also obtain evidences of the TDS bands in Fe(Te,Se) by measuring the associated surface states. Although they generally overlap with bulk states on the (001) surface, their spin-polarized character provides a unique signature to those surface states, detectable via spin-resolved photoemission measurements, as already shown in Fig.3. 
The intensity plots of Fe(Te,Se) from SARPES are displayed in Fig.~5(a - b). The spin-integrated plot is the same as the one in Fig.~4(a), showing clearly the hybridization of $d_{xy}$ and $d_{yz}$ obitals, while the spin-resolved intensity plot (the intensity difference between spin-up and spin-down photoelectrons) shows spin-polarization of the $d_{yz}$ band near $E_{\text{F}}$. As shown in Fig.~5(c) inset, we measured four cuts, and all the four spin-resolved EDCs [Fig.~5(c - f)] show clear spin-polarizations [Fig.~5(g - h)], exhibiting a helical texture. To further confirm that the spin-polarizations are the intrinsic properties of the electronic states in the crystal and not induced by the photoelectron process, we double-checked the spin-polarization with different photon energies in a synchrotron facility, as shown in Fig.~5(i) and with different photon polarizations, as shown in Supplementary Information Part VI. All the results consistently show a spin-helical texture, excluding the possibility that the spin-polarizations come from the photoelectron process or spin matrix-element effect. The magnitude of spin-polarization of the $d_{yz}$ band is about 50\%, indicating a coexistence of unpolarized bulk and polarized surface states. The results are consistent with the spin-resolved spectrum from calculations, as shown in Fig.~5(j).

\begin{figure}[!htb]
\begin{center}
\includegraphics[width=.5\textwidth]{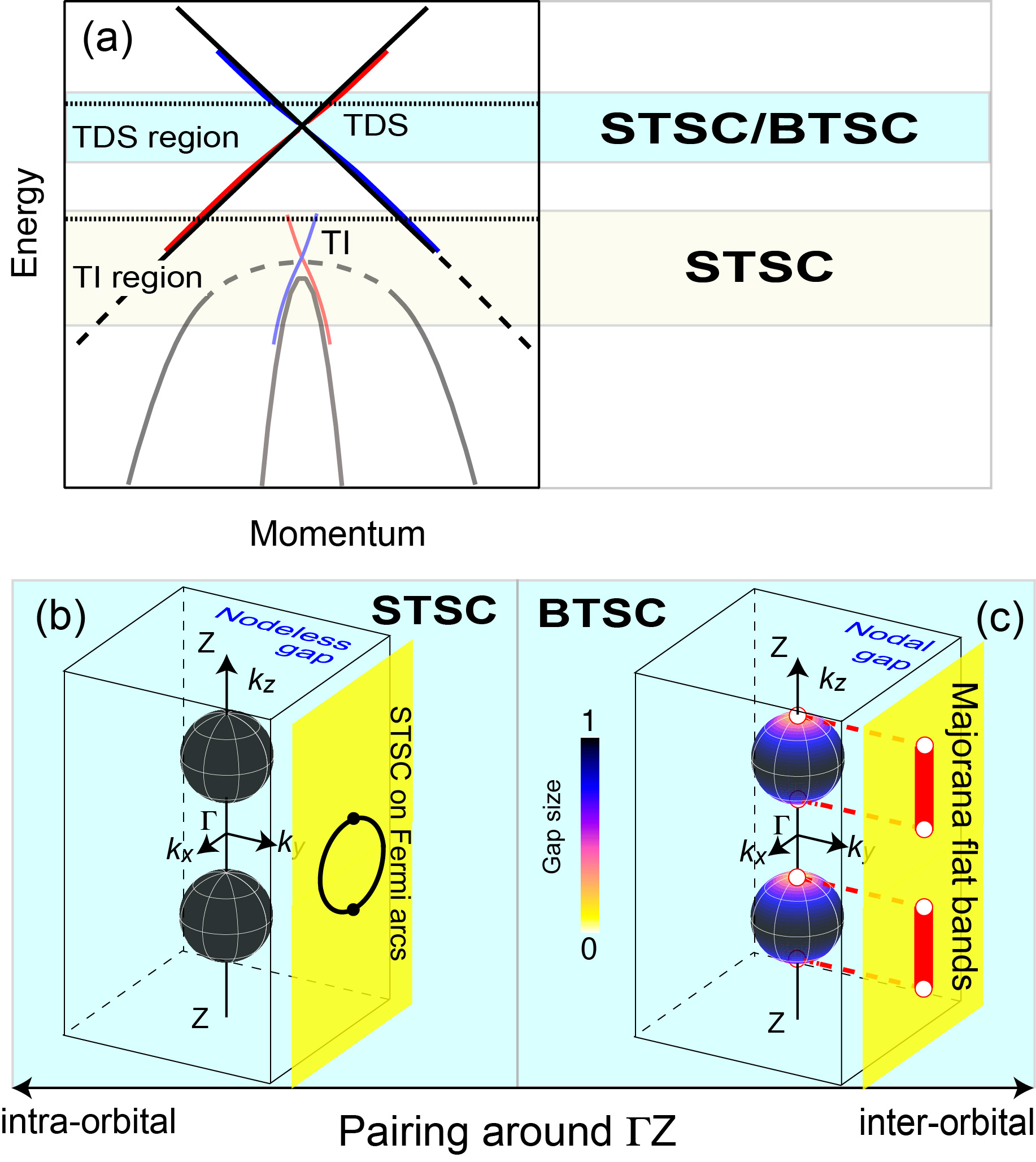}
\end{center}
 \caption{\label{theory} \textbf{Topological states and the related topological superconductivity.} (a) The in-plane band structure of Fe(Te,Se) at the bulk Dirac point. There will be surface topological superconductivity (STSC) or bulk topological superconductivity (BTSC) if $E_{\text{F}}$ locates in the corresponding region. In the case that $E_{\text{F}}$ locates in the TDS region, there will be two spherical FSs along $\Gamma$Z, and (b) if the intra-orbital pairing around $\Gamma$Z dominates, the SC gap will be nodeless and the Fermi arcs on side surfaces will be topologically superconducting; (c) if the inter-orbital pairing around $\Gamma$Z dominates, the SC gap will be isotropic in plane and has nodes along $k_z$. In this case, there is topological superconductivity in bulk and Majorana fermions on side surfaces. }
\end{figure}

The coexistence of TI and TDS bands at different Fermi levels in iron-based superconductors provides a basis for a rich variety of possible topologically superconducting states. Based on the band structure of Fe(Te,Se), the right part of Fig.~6(a) illustrates the possible superconducting states as one were to shift $E_{\text{F}}$ to the TI or TDS Fermi level region via charge carrier doping. In both the TI and TDS regions, spin-helical FSs of surface states are to be expected. For the TDS, the spin-helical FSs (Fermi arc pairs) appear on some side surfaces~\cite{FangPRB2012, HasanScience2015}, with two spherical bulk FSs along the $\Gamma$Z line, as shown in Fig.~6(b). Invoking the notion that spin-helical surface states in proximity to a bulk $s$-wave superconductor feature topologically superconducting states with MBSs in its associated vortex cores~\cite{FuPRL2008}, the surface states in the TI region are expected to display topological superconductivity, which is already observed by ARPES and STM measurements~\cite{ZhangScience2018, DingSTM}. Similarly, if bulk $s$-wave pairing persists in the TDS region, the spin-helical FSs on side surfaces are likewise expected to form topologically superconducting states [Fig.~6(b)]. 
However, since $d$ orbitals dominate the Fermi level density of states in iron-based superconductors and exhibit strong correlation effects such as Hund's coupling, the inter-orbital pairing may dominate in the TDS Fermi level region, which would generate a spin triplet pairing state on the two spherical bulk FSs with point nodes on $k_z$ axis [Fig.~6(c)], as a consequence of orbit-momentum locking in the bulk Dirac cone~\cite{SatoPRL2015, SatoPRB2016}. Such a scenario would hence yield yet another intriguing pairing state, namely a bulk topological superconductor, which would host Majorana fermions on its side surfaces. 

A shift of the Fermi level could also induce a topological phase transition from topological-superconductor to trivial-superconductor, as predicted in Ref.~\citenum{XuPRL2016}, which might be useful to optimize the conditions for surface topological superconductivity and MBSs. It is, however, difficult to study such proposals in the scope of Fe(Te,Se), as it appears tedious to change the electron doping of Fe(Te,Se) without significantly affecting the coherence of the electronic states. Fortunately, the Fermi level of Li(Fe,Co)As is easily tuned by Co content, and the TI/TDS bands may be separately accessed, rendering Li(Fe,Co)As ideal for such studies. One important aspect to address is the overlapping character between the trivial bulk states and the topological bands, which might interfere with the topological pairing, or might stabilize topological superconductivity over a larger range of doping\cite{HughesPRL2012}. We defer a more detailed discussion of these issues to future work.

Our findings of the TI and TDS states in Li(Fe,Co)As and Fe(Te,Se) prove the generic existence of topological bands in iron-based superconductors. Their simple structures, multiple topological states, and a tunable Fermi level make iron-based superconductors ideal platforms for the study of topological superconductivity, MBSs, and as such, potentially, topological quantum computation.

\textbf{Methods}

High quality single crystals of LiFe$_{1-x}$Co$_x$As were synthesized by self-flux method. The Fe(Te,Se) single crystals of sample \#1 were grown by the self-flux method. The composition is Fe$_{1+y}$Te$_{0.57}$Se$_{0.43}$, with $y$ = 14\%, as detected by the inductively coupled plasma (ICP) atomic emission spectroscopy. The as-grown Fe(Te,Se) single crystals were annealed in controlled amount of O$_2$ to remove excess Fe~\cite{SunST2013}. The Fe(Te,Se) single crystals of sample \#2 were grown by the Bridgman technique, with a composition of FeTe$_{0.55}$Se$_{0.45}$. The as-grown single crystals contain no or very small amount of excess Fe. No annealing process was applied to sample \#2~\cite{WenRPP2011}. Both Fe(Te,Se) samples show a $T_\mathrm{c}$ of $\sim$ 14.5 K and the same band structure.

The high resolution ARPES measurements on Li(Fe,Co)As were performed on a spectrometer with a VG-Scienta R4000WAL electron analyzer. The energy resolution of the system was set to $\sim$ 5 meV. The spin-resolved ARPES (SARPES) measurements on Li(Fe,Co)As were carried out with a ScientaOmicron DA30-L analyzer, together with twin very-low-energy- electron-diffraction (VLEED) spin detectors~\cite{YajiRSI2016}. The energy resolution for the spin-resolved mode was set to $\sim$ 6 meV for $x = 3\%$ and 9\% samples, and $\sim$ 12 meV for $x = 12\%$ sample. All the ARPES measurements on Li(Fe,Co)As are carried out with a 6.994-eV laser.
The same laser SARPES system is used for the high-resolution measurements on Fe(Te,Se). The photon-dependent SARPES measurements on Fe(Te,Se) were carried out at BL9B, HiSOR. The resolution is set to $\sim$ 6 meV for the laser SARPES, and $\sim$ 30 meV for the SARPES at HiSOR.

The measurements on Fe(Te,Se) of the in-plane magnetoresistance $\rho (H)$ in static magnetic field up to 14 T were carried out with a commercial Physical Property Measurement System (PPMS). The measurements in pulsed high magnetic fields up to 30 T were performed with a four-probe point contact method. The experimental data taken with pulsed magnetic field were recorded on a 16 bit digitizer and were analyzed using a numerical lock-in technique.

Our Density functional theory calculations employ the projector augmented
wave  method encoded in Vienna \textit{ab initio} simulation
package\cite{Kresse1993,Kresse1996,Kresse1996prb}, and the local density approximation for the exchange correlation functional is used\cite{Perdew1996prl}.
Throughout this work, the cutoff energy of 500 eV is taken for expanding the wave functions into plane-wave basis. In the
calculation, the Brillouin zone is sampled in the $k$ space within
Monkhorst-Pack scheme\cite{Monkhorst1976}. The number of these $k$ points depends on materials: 11 $\times$ 11 $\times$  5 and 9$\times$ 9 $\times$ 9 for LaOFeAs, LiFeAs, Fe(Te,Se) conventional cells and BaFe$_2$As$_2$ primitive cell, respectively. The spin-orbit coupling (SOC) was included in the self-consistent calculations of electronic structure.

The effective Hamiltonian for the theoretical calculations on Fe(Te,Se) was built on the eight bands ($p_z$, $d_{xy}$, $d_{yz}$/$d_{xz}$) at the $\Gamma$ point. At first, we derived the 4-band model without SOC. The first-principles calculations indicate the four characteristic bands are labeled as the irreducible representations $\Gamma^-_2$, $\Gamma^+_5$, and $\Gamma^+_4$ of the point-group $D_{4h}$ at $\Gamma$ without SOC. The 4-band time-reversal-invariant $k\cdot p$ model was established under the basis of those irreducible representations. Then, the SOC was taken into consideration, by doubling the basis with the spin degree of freedom and introducing additional terms. The (001) surface spectrum is computed by the surface Green's function method. (More detail can be found in Supplementary Information Part III).

\begin{addendum}
\item We acknowledge K. Asakawa, A. Harasawa, Y. Hesagawa, D. Hirai, Z. Hiroi, K. Ishizaka, N. Mitsuishi, M. Sakano, Y. Yoshida for experimental assistance. This work was supported by the Photon and Quantum Basic Research Coordinated Development Program from MEXT, JSPS (KAKENHI Grant Nos. 25220707, JP17H02922, JP16K17755 and 17H01141), and the Grants-in-Aid for Scientific Research on Innovative Areas ``Topological Material Science'', JSPS (Grant No. JP15H05855 ). The work in Brookhaven is supported by the Office of Science, U.S. Department of Energy under Contract No. DE-SC0012704 and the Center for Emergent Superconductivity, an Energy Frontier Research Center funded by the U.S. Department of Energy, Office of Science. The work in W\"urzburg is supported by ERC-StG-TOPOLECTRICS-336012, DFG-SFB 1170, and DFG-SPP 1666.

\item[Competing Interests] The authors declare that they have no competing financial interests.
\item[Correspondence] Correspondence and request for materials should be addressed to P.Z. or S.S. (emails: zhangpeng@issp.u-tokyo.ac.jp, shin@issp.u-tokyo.ac.jp)
\item[Author contributions] P.Z. performed the ARPES measurements on Li(Fe,Co)As and analyzed the data with help from K.Y., T.Kondo and S.S.. X.Wu, J.H. and R.T. performed the DFT calculations. G.D., X.Wang and C.J. synthesized the Li(Fe,Co)As samples. P.Z. performed the ARPES measurements on Fe(Te,Se) and analyzed the data with help from Y.I., K.Y., C.B., K.Kuroda, T.Kondo, K.O., K.S., S.W., K.M., T.O., H.D. and S.S..  P.Z., Y.K. and K.Kindo performed the MR measurements on Fe(Te,Se). Z.W., X.Wu, R.T., T.Kawakami and M.S. performed the theory calculations on Fe(Te,Se). G.D.G., Y.S. and T.T. synthesized the Fe(Te,Se) samples. All authors discussed the manuscript. P.Z. and S.S. supervised the whole project.
\end{addendum}

\end{document}